\documentstyle[12pt,epsf]{article}

\newcommand{\gev}{\ {\rm GeV}}
\newcommand{\mev}{\ {\rm MeV}}
\newcommand{\lan}{{\langle}}
\newcommand{\ran}{{\rangle}}

\newcommand{\vspu}{\vspace*{5mm}}

\newcommand{\prd}{Phys. Rev. D }
\newcommand{\prl}{Phys. Rev. Lett. }
\newcommand{\plb}{Phys. Lett. B }
\newcommand{\npb}{Nucl. Phys. B }
\newcommand{\zpc}{Z. Phys. C }
\newcommand{\epjc}{Eur. Phys. J. C }

\newcommand\be{\begin{equation}}
\newcommand\ee{\end{equation}}
\newcommand\bea{\begin{eqnarray}}
\newcommand\eea{\end{eqnarray}}

\makeatletter \@addtoreset{equation}{section} \makeatother

\begin{document}
\baselineskip 20pt
\begin{center}
{\large\bf Strong Couplings of Heavy Mesons to A Light Vector Meson in QCD}
\vspu \\
\ \\
Zuo-Hong Li$^{\rm a,b,c,d,}$\footnote{Email: lizh@ytu.edu.cn}, Tao Huang$^{a,c}$
Jin-Zuo Sun$^{a,b}$ and Zhen-Hong Dai$^{b}$\\
\vspu \ \\
\footnotesize{a. CCAST (World Laboratory), P.O.Box 8730, Beijing 100080, China\\
\vspu b. Department of Physics, Yantai University, Yantai 264005,
China\footnote{Mailing address}\\\vspu c. Institute of High Energy Physics, P.O.Box
918(4), Beijing 100039, China \\\vspu d. Department of Physics, Peking University,
Beijing 100871, China
\\
}
\end{center}
\vspu
\begin{center}{ \bf ABSTRACT}
\end{center}
\normalsize

We make a detailed analysis of the $BB\rho$($DD\rho$) and $B^*B\rho$($D^{\ast}D\rho$)
strong couplings $g_{BB\rho}$($g_{DD\rho}$) and $g_{B^*B\rho}$($g_{D^{\ast}D\rho}$)
using QCD light cone sum rules(LCSR). The existing some negligence is pointed out in
the previous LCSR calculation on $g_{B^*B\rho}$($g_{D^{\ast}D\rho}$) and an updated
estimate is presented. Our findings can be used to understand the behavior of the $B,\
D\to \rho$ semileptonic form factors at large momentum transitions. 

\newpage \baselineskip 20pt
\begin{center}
{\bf{\large \bf 1. INTRODUCTION}}
\end{center}

At the present time, there is an increasing interest in exclusive
$B$ decays in order to explore the sources of CP violation.
However, the definite interpretations for the relevant
experimental data demand that we have the ability to precisely
compute the physical amplitudes. It is nonperturbative QCD
dynamics not being dealt rigorously with that would hinder us from
doing such a desired calculation. One is forced to use some
approximate methods. Lattice QCD simulation is the most
trustworthy approach to nonperturbative QCD effects, with no
parameters or assumptions, but the precision of calculation is
limited by the available computing resources and some certain
restriction exists\cite{1}, for example, in describing
$b\rightarrow u$ or $s$ transitions. The formulation of QCD
factorization formula\cite{2} is based on the first principle and
is viewed as a great progress in phenomenology of heavy flavors;
however, the underlying long distance effects included in a series
of the hadronic matrix elements still confront us. Heavy quark
symmetry can very well apply to describing the systems including
one heavy quark, but is of less predictive power for
heavy-to-light transitions. Although QCD sum rule method\cite{3}
builds its underlying physical assumptions on the field theory,
and exhibits its decided superiority in dealing with some of
nonperturbative quantities, such as decay constants, hadronic
matrix elements and strong coupling constants, a problem with it
is that the resulting form factors for heavy-to-light transitions
can not behave very well in the heavy quark limit
$m_Q\rightarrow\infty$. The successes of perturbative QCD in
treating numerous hard exclusive processes have excited the
occurrence of many excellent works. The most prominent of them is
the development of QCD light-cone sum rules (LCSR)\cite{4,5}. It
is the striking advantage of this approach to describe
heavy-to-light transitions in a way consistent with the
universally accepted physical picture that nonperturbative QCD
dynamics occupies an dominant place and perturbative hard gluon
exchanges contribute only a subleading effect in that case. LCSR
approach takes the basic correlator, in which the proper current
operators are sandwiched between the vacuum and an on shell light
meson state, and adopts the operator product expansion (OPE)
around the light cone $x^{2}\approx 0$ instead of at the small
distance $x\approx 0$. Nonlocal matrix elements occurring in this
approach, which encode all the information on large distance
dynamics, are parametrized in terms of a set of so-called
light-cone wavefunctions classified by twist, which describe the
momentum distributions of the quarks inside the relevant light
mesons. This is identical with a summation over all the condensate
terms in the short distance OPE. Consequently, the resulting form
factors for heavy-to-light transitions exhibit the correct
behavior with heavy quark mass, averting the problem with the
traditional sum rules. This method has extensively been accepted
and has found its successful applications, such as the
investigations of the form factors for heavy-to-light transitions
at small and intermediate momentum transfers[5-11] and of the
strong couplings between heavy and light mesons\cite{8,12,13},
since its presentation in\cite{4}. Very recently, it has been
generalized to study the nonfactorizable effects in $B\rightarrow
\pi\pi$\cite{14} and to probe heavy-to-light form factors in the
whole kinematically accessible ranges\cite{15}. The technical
details of LCSR can be found, for instance, in \cite{6}, while for
a detailed comparison with traditional sum rules, see \cite{9}.

An investigation of strong interactions between heavy mesons and a light vector meson
is of important phenomenological interest, and especially the relevant strong
couplings can be employed to understand the behavior of the corresponding
heavy-to-light form factors at large momentum transfer in a pole dominance model,
which has proven to be exact in an effective chiral Lagrangian approach\cite{16}. The
off-shell $B^{\ast}B\rho$($D^{\ast}D\rho$) strong coupling
$g_{B^{\ast}B\rho}$($g_{D^{\ast}D\rho}$), in fact, has been calculated in the standard
LCSR approach\cite{13}; however, it calls for a reexamination due to some negligence
existing in calculations. Another nonperturbative quantity deserving of investigation
is the $BB\rho$($DD\rho$) strong coupling $g_{BB\rho}$($g_{DD\rho}$), which turns out
to be equally important. Motivated by all these facts, in this paper we make a
systematic study on them.

This presentation is organized as follows. The following Section
is devoted to a detailed derivation of the sum rules for
$g_{BB\rho}$($g_{DD\rho}$) and
$g_{B^{\ast}B\rho}$($g_{D^{\ast}D\rho}$), using LCSR method. Then
we give a numerical analysis of the resulting sum rules, including
a discussion of error estimates, in Sec.3. The last Section give
to a simple summary.
\begin{center}
\noindent{\large \bf 2. LCSR'S FOR THE STRONG COUPLINGS }
\end{center}

The $BB\rho$ and $B^{\ast}B\rho$ strong couplings $g_{BB\rho }$ and $g_{B^{\ast }B\rho
}$ can be defined as
\begin{equation}
\left\langle \rho \left( q,e\right) B\left( p\right) |B\left( p+q\right) \right\rangle
=g_{BB\rho }e_{(\lambda)}^{\ast }\cdot p,  \label{1}
\end{equation}
\begin{equation}
\left\langle \rho \left( q,e\right) B^{\ast}(p,\eta)|B\left( p+q\right) \right\rangle
=-g_{B^{\ast}B\rho }\epsilon_{\mu\alpha\beta\gamma}\eta^{\ast\mu}q^{\alpha} e^{\ast
\beta}_{(\lambda)}p^{\gamma}. \label{2}
\end{equation}%
The resulting findings can easily converted into the corresponding $c$-quark meson
cases. According to the general strategy of QCD sum rules, it is needed to construct
an adequate correlator in order to obtain an qualitative estimate for $g_{BB\rho }$
and $g_{B^*B\rho }$. As usual, we use two correlators of the following forms as the
starting points of LCSR calculations on $g_{BB\rho }$ and
$g_{B^*B\rho }$, respectively,%
\begin{eqnarray}
F\left( p,\left( p+q\right) \right) &=&i\int d^{4}xe^{ip\cdot x}\left\langle \rho
\left( q,e\right) \left| T\overline{u}\left( x\right) i\gamma _{5} b\left( x\right)
,\overline{b}\left( 0\right) i\gamma
_{5}d\left( 0\right) \right| 0\right\rangle  \nonumber \\
&=&\widetilde{F}\left( p^{2},\left( p+q\right) ^{2}\right) e^{\ast }\cdot p, \label{3}
\end{eqnarray}
\begin{eqnarray}
G_{\mu}(p,p+q)&=&i\int d^4x e^{ip\cdot x}
\left\langle\rho(q,e)|T\overline{u}(x)\gamma_{\mu}b(x),\overline
b{}(0)i\gamma_5d(0)|0\right\rangle \nonumber\\
&=&\widetilde{G}\left( p^2,\left( p+q\right)^2
\right)\epsilon_{\mu\alpha\beta\gamma}q^{\alpha}e^{\ast\beta}p^{\gamma}.\label{4}
\end{eqnarray}

For the correlator (3), isolating the pole contribution of the lowest $0^{-}\ B$ meson
and parametrizing these from the higher $0^{-}$ states in a form of dispersion
integral, the hadronic form of the invariant function $\tilde{F}\left( p^2,\left(
p+q\right)^2 \right) $ may be written as
\begin{eqnarray}
\tilde{F}^H\left( p^2,\left( p+q\right)^2 \right) &=&\frac{
m_{B}^{4}/m_{b}^{2}f_{B}^{2} g_{BB\rho }}{\left(m_{B}^{2}-
p^{2}\right) \left[ \left(m_{B}^{2}- p+q\right) ^{2}\right] }+\int
\int \frac{\rho _1^{H}\left( s_{1},s_{2}\right) }{\left(
s_{1}-p^{2}\right) \left[ s_{2}-\left( p+q\right) ^{2}\right]
}ds_{1}ds_{2}, \label{5}
\end{eqnarray}%
with a double integral starting from the same threshold parameter
$s_{0}$, which should be set in the neighborhood of the squared
mass of the first excited $ 0^{-}\ B$ meson. Similarly we have for
the invariant function $\tilde{G}^{H}\left( p,\left( p+q\right)
\right)$,
\begin{equation}
\tilde{G}^{H}\left(p^2,\left(p+q\right)^2
\right)=\frac{m_{B^*}m_B^2f_Bf_{B^*}g_{B^*B\rho}}
{m_b\left(m_{B^*}^2-p^2\right)\left[m_B^2-(p+q)^2\right]}+\int
\int
\frac{\rho_2^H\left(s_1,s_2\right)ds_1ds_2}{\left(s_1-p^2\right)\left[s_2-(p+q)^2\right]}.\label{6}
\end{equation}

QCD calculations of the underlying correlators may be allowed, on the other side, for
the negative and large values of $p^{2}$ and $ \left( p+q\right) ^{2}$, in which case
the $b$ quarks travel only a small distance $x$ and therefore the operator product
expansion (OPE) goes effectively in powers of the deviation from the light cone
$x^{2}\approx 0.$ We would like to work in the case where the interactions of the $b$
quarks with the background field gluons are omitted, since their influence on the sum
rules is, as always, negligibly small. On contracting the $b$ quark operators into a
free propagator,
\begin{equation}
\lan 0|Tb\left( x\right) \overline{b}\left( 0\right)|\ran =\frac{1}{\left( 2\pi
\right) ^{4}i}\int d^{4}ke^{-ik\cdot x}\frac{k\!\!\!/+m_{b}}{m_{b}^{2}-k^{2}},
\label{7}
\end{equation}
one gets,
\begin{eqnarray}
\widetilde{F}^{QCD}\left(p^2,\left(p+q\right)^2 \right)
=&&\frac{1}{(2\pi)^4}\int \int d^{4}x d^{4}k\frac{1}{
m_{b}^{2}-k^{2}}e^{i(p-k)\cdot x}\left[k^{\mu }\left\langle \rho
\left( q,e\right) \left| T\overline{u}\left( x\right) \gamma _{\mu
}d\left( 0\right)
\right| 0\right\rangle\right.\nonumber\\
&&\left.-m_b\left\langle \rho (q,e)\left| T\overline{u}\left( x\right) d\left(
0\right) \right| 0\right\rangle\right]. \label{8}
\end{eqnarray}
The nonlocal matrix elements $\left\langle \rho \left( q,e\right) \left| T\overline
{u}\left ( x\right) \gamma _{\mu }d\left( 0\right) \right| 0\right\rangle $ and
$\left\langle \rho (q,e)\left| T\overline{u}\left( x\right) d\left( 0\right) \right|
0\right\rangle$ define the light cone wavefunctions of the $\rho $\ meson[17-19] as,
\begin{eqnarray}
\left\langle \rho \left( q,e\right) \left|\overline{u}\left( x\right) \gamma _{\mu
}d\left( 0\right) \right| 0\right\rangle &=&f_{\rho }m_{\rho }\left\{ \frac{e^{\left(
\lambda \right) \ast }\cdot x}{q\cdot x}q_{\mu }\int_{0}^{1}due^{iuq\cdot x}\left[
\varphi _{\parallel }\left( u,\mu \right) +\frac{m_{\rho }^{2}x^{2}}{16}A\left( u,\mu
\right) \right] \right.
\nonumber \\
&&+\left( e_{\mu }^{\left( \lambda \right)\ast }-q_{\mu }\frac{e^{\left( \lambda
\right) \ast }\cdot x}{q\cdot x}\right) \int_{0}^{1}due^{iuq\cdot x}g_{\perp
}^{\left( v\right) }\left( u,\mu \right)  \nonumber \\
&&\left. -\frac{1}{2}x_{\mu }\frac{e^{\left( \lambda \right)\ast }\cdot x}{%
\left( p\cdot x\right) ^{2}}m_{\rho }^{2}\int_{0}^{1}due^{i\xi \cdot x}C\left( u,\mu
\right) \right\},  \label{9}
\end{eqnarray}%
\begin{equation}
\left\langle \rho (q,e)\left| T\overline{u}\left( x\right) d\left( 0\right) \right|
0\right\rangle=-i/2f^T_{\rho}m^2_{\rho}(e^{\lambda}\cdot x)\int^1_0 du e^{iuq\cdot
x}h^s_{\parallel}(u,\mu_b).      \label{10}
\end{equation}
$f_{\rho}$ stands for the usual vector decay constant of the $\rho$ meson and
$f_{\rho}^T$ is defined as $\left\langle0|\bar{d}\sigma_{\mu\nu}u|
\rho\right\rangle=if_{\rho}^T\left(e_{\mu}^{ (\lambda)}q_{\nu}-e_{\nu}^{(\lambda)}
q_{\mu}\right)$; $\varphi _{\parallel }\left( u,\mu \right) $ is the leading twist-2
wavefunction, $g_{\perp }^{\left( v\right) }\left( u,\mu \right) $ and
$h^s_{\parallel}(u,\mu_b)$ refer to the twist-3 ones, and both $A\left( u,\mu \right)
$ and $C\left( u,\mu \right) $ have twist-4, which parametrize the mass corrections.
$\varphi _{\parallel }\left( u,\mu \right) $, $g_{\perp }^{\left( v\right) }\left(
u,\mu \right) $ and $h^s_{\parallel}(u,\mu_b)$ are normalized as $\int_{0}^{1}duf
\left( u\right) =1$, while $C\left( u,\mu \right) $ satisfies $\int_{0}^{1}duC\left(
u\right) =0$. A tedious but straightforward calculation yields
\begin{eqnarray}
\widetilde{F}^{QCD}\left( p^{2},\left( p+q\right) ^{2}\right)
&=&f_{\rho }m_{\rho }\left[ \int_{0}^{1}du \frac{\varphi
_{\parallel }\left( u\right) }{m_{b}^{2}-\left(
p+uq\right)^{2}}-\frac{1}{m_{b}^{2}-\left( p+q\right) ^{2}}\right]  \nonumber \\
&&+f_{\rho}^Tm_{\rho}^2m_b\int_0^1du\frac{h^{(s)}_{\parallel}(u,\mu_b)}{\left[
m_{b}^{2}-\left( p+uq\right)^{2}\right] ^{2}} \nonumber \\
&&-\frac{1}{2}f_{\rho }m_{\rho }^{3}\int_{0}^{1}du\left\{\frac{1}{2\left[
m_{b}^{2}-\left( p+uq\right)^{2}\right] ^{2}} + \frac{m_b^2}{\left[ m_{b}^{2}-\left(
p+uq\right)^{2}\right]^{3}}\right\}. \nonumber\\
&&\times\left[ A\left( u\right)+8\widetilde{C}\left( u\right) \right] \label{11}
\end{eqnarray}%
In deriving Eq.(11), it proves to be convenient for a partial
integration to introduce the auxiliary functions
$\overline{f}\left( u\right)=\int_{0}^{u}f\left( v\right) dv$ for
$f(u)=\varphi_{\parallel }(u,\mu)$, $g_{\perp}^{(v)}(u,\mu)$ and
$A(u,\mu)$, and $\overline{C}(u)=\int _0^uC(v)dv$,
$\widetilde{C}\left( u\right) =\int_{0}^{u} \overline{C} \left(
v\right) dv$, which is equal to zero when $u=1$. The twist-3
contribution from $ g^{(v)}_{\perp}{(u,\mu)}$ vanishes exactly due
to cancellations in the partial integrations.

Furthermore, it is indispensable to convert Eq.(11) into a form of dispersion integral
for upcoming continuum substraction. The relevant QCD spectral density can easily be
obtained by virtue of the technique suggested in \cite{20}. In the following we
consider only the twist-2 and -3 terms and give a detailed deviation. First of all, we
perform a double Borel transformation $Q_{1}^{2}=-p^{2}\rightarrow
M_{1}^{2},Q_{2}^{2}=-\left( p+q\right) ^{2}\rightarrow M_{2}^{2},$ for the twist-2
term ( the term proportional to the inverse $m_{b}^{2}-\left( p+q\right) ^{2}$
disappears after doing that ), obtaining
\begin{eqnarray}
\widehat{B}\left( M_{1}^{2},Q_{1}^{2}\right) \widehat{B}\left(M_{2}^{2},
Q_{2}^{2}\right) \int_{0}^{1}du\frac{\varphi_{\parallel}\left( u\right)
}{m_{b}^{2}-\left( p+uq\right) ^{2}}=\frac{M^{2}}{M_{1}^{2}M_{2}^{2}}e^{-\frac{1}
{M^{2}}\left[ m_{b}^{2}+m_{\rho }^{2}u_{0}\left( 1-u_{0}\right) \right]
}\varphi_{\parallel}\left( u_{0}\right) , \label{12}
\end{eqnarray}%
with $u_{0}=M_{1}^{2}/\left( M_{1}^{2}+M_{2}^{2}\right) $ and $M^{2}=\frac{
M_{1}^{2}M_{2}^{2}}{M_{1}^{2}+M_{2}^{2}}.$ The symmetry of the correlator makes it
natural to set $M_{1}^{2}=M_{2}^{2}$ so that the wavefunctions $ \varphi_{\parallel}
\left( u\right) $ may take its value at the symmetric point $u_{0}=1/2$. Further,
making a replacement $M_{1}^{2}\rightarrow 1/\sigma _{1},M_{2}^{2}\rightarrow 1/\sigma
_{2}$ in Eq.(12) yields
\begin{eqnarray}
\widehat{B}\left( \frac{1}{\sigma _{1}},Q_{1}^{2}\right) \widehat{B}\left( \frac{1}
{\sigma _{2}},Q_{2}^{2}\right) \int_{0}^{1}du\frac{\varphi_{\parallel}\left( u\right)
}{m_{b}^{2}-\left( p+uq\right) ^{2}} &=&\frac{\sigma _{1}\sigma _{2}}{ \sigma
_{1}+\sigma _{2}}e^{-\left( \sigma _{1}+\sigma _{2}\right) \left( m_{b}^{2}
+\frac{1}{4}m_{\rho}^{2}\right) }  \nonumber \\
&=&f\left( \sigma _{1},\sigma _{2}\right).  \label{13}
\end{eqnarray}%
Finally, we take the function $\overline{f}\left( \sigma _{1},\sigma _{2}\right)=
\frac{1}{ \sigma _{1}\sigma _{2}}f\left( \sigma _{1},\sigma _{2}\right)$ and perform
one more Borel transformation in the variables $\sigma _{1}$ and $\sigma _{2}$:
$\sigma_1\rightarrow 1/s_1$ and $\sigma_2\rightarrow 1/s_2$. The resulting QCD
spectral density reads
\begin{eqnarray}
\rho_{tw2} ^{\left( QCD\right) }\left( s_{1},s_{2}\right) &=&\frac{f_{\rho}m_{\rho}}
{s_{1}s_{2}} \widehat{B}\left( \frac{1}{s_{1}},\sigma _{1}\right) \widehat{B}\left(
\frac{ 1}{s_{2}},\sigma _{2}\right) \overline{f}\left(
\sigma _{1},\sigma_{2}\right)  \nonumber \\
&=&f_{\rho}m_{\rho}\varphi _{\parallel }\left( \frac{1}{2} \right) \delta \left(
s_{1}-s_{2}\right)\Theta \left( s_{1}-m_{b}^{2}-\frac{1}{4}m_{\rho }^{2}\right) \Theta
\left( s_{2}-m_{b}^{2}-\frac{1}{4}m_{\rho }^{2}\right).  \label{14}
\end{eqnarray}%
Applying all the same procedure to the twist-3 part, we have
\begin{eqnarray}
\rho_{tw3} ^{\left( QCD\right) }\left( s_{1},s_{2}\right)
&=&f_{\rho}^Tm_{\rho}^2m_b\left( m_{b}^{2}+\frac{1}{4}m_{\rho
}^{2}\right)^2h^s_{\parallel}\left(\frac{1}{2}\right)\delta \left(
s_{1}-m_{b}^{2}-\frac{1}{4}m_{\rho }^{2}\right) \delta \left(
s_{2}-m_{b}^{2}-\frac{1}{4}m_{\rho }^{2}\right)\frac{1}{s_{1}s_{2}}, \nonumber\\
&&\ \label{15}
\end{eqnarray}
which means that the twist-3 part receives no continuum
substraction. With Eqs. (14) and (15), $ \widetilde{F}^{QCD}\left(
p^{2},\left( p+q\right) ^{2}\right)$ can be expressed as
\begin{eqnarray}
\widetilde{F}^{QCD}\left( p^{2},\left( p+q\right) ^{2}\right)
&=&f_{\rho }m_{\rho }\int \int ds_{1}ds_{2}\frac{ \delta \left(
s_{1}-s_{2}\right)  \Theta \left( s_{1}-m_{b}
^{2}-\frac{1}{4}m_{\rho }^{2}\right) \Theta \left(
s_{2}-m_{b}^{2}-\frac{1}{4}m_{\rho }^{2}\right) }{ \left(s_{1}-
p^{2}\right) \left[s_{2}- \left( p+q\right) ^{2}\right] }\varphi
_{\parallel }\left( \frac{1}{2}\right)
\nonumber \\
&&+ \frac{f_{\rho }^Tm_{\rho }^{2}m_bh^s_{\parallel}\left(\frac{1}{2}\right)
}{\left(m_{b}^{2} +\frac{1} {4}m_{\rho
}^{2}-p^2\right)\left(m_{b}^{2}+\frac{1}{4}m_{\rho }^{2}-(p+q)^2\right)}
\nonumber \\
&&-\frac{1}{2}f_{\rho }m_{\rho }^{3}\int_{0}^{1}du\left\{\frac{1}{2\left[
m_{b}^{2}-\left( p+uq\right)^{2}\right] ^{2}} + \frac{m_b^2}{\left[ m_{b}^{2}-\left(
p+uq\right) ^{2}\right]^{3}}\right\} \left[ A\left( u\right) +8\widetilde{C}\left(
u\right)\right] .\nonumber\\
&&\              \label{16}
\end{eqnarray}

Making the Borel improvement $p^2\rightarrow M_1^2$, $(p+q)^2\rightarrow M^2_2$ for
both the theoretical and hadronic expressions, which suppresses the higher state and
twist-4 contributions, and then equating them by the use of the quark-hadron duality
ansatz, the final sum rule for $f_{B}^{2}g_{BB\rho }$ reads,
\begin{eqnarray}
f_{B}^{2}g_{BB\rho }&=&\frac{m_{b}^{2}}{m_{B}^{4}}e^{\frac{m_{B}^{2}}{M^{2}}}
\left\{f_{\rho}m_{\rho}M^2\left[e^{-\frac{1}{M^2}
\left(m_b^2+\frac{1}{4}m_{\rho}^2\right)}-e^{-\frac{s_0}{M^2}}\right]
\varphi_{\parallel}\left(\frac{1}{2}\right)+\frac{1}{4}m_{\rho
}^{2}e^{-\frac{1}{M^{2}}\left( m_{b}^{2}+\frac{1}{4}m_{\rho }^{2}\right)}\right.\nonumber\\
&&\left.\times\left[4m_{b}f_{\rho }^{T}h_{\parallel }^{\left(
s\right)}\left(\frac{1}{2}\right)-f_{\rho }m_{\rho }\left[ A\left(
\frac{1}{2}\right)+8 \widetilde{C}\left( \frac{1}{2}\right) \right] \left(
1+\frac{m_{b}^2}{M^{2}}\right) \right] \right\}. \label{17}
\end{eqnarray}

Now let's turn from this topic to a discussion of the sum rule for
the $B^{\ast }B\rho $ strong coupling $g_{B^{\ast }B\rho }$ by
expanding the relevant correlator (4) around the light cone
$x^{2}=0$. Utilizing Eq.(7) it follows immediately that
\begin{eqnarray}
G_{\mu }^{QCD}(p^2,(p+q)^2)=&& \frac{i}{ \left( 2\pi \right)
^{4}}\int \int d^{4}xd^{4}k\frac{1}{m_{b}^{2}-k^{2}} e^{i\left(
p-k\right)\cdot x}\left[k^{\nu }\left\langle \rho \left|
T\overline{u}\left( x\right) \gamma _{\mu }\gamma _{\nu }\gamma
_{5}d\left( 0\right)
\right| 0\right\rangle \right.\nonumber\\
&&\left.+m_b\left\langle \rho \left| T\overline{u}\left( x\right) \gamma _{\mu }\gamma
_{5}d\left( 0\right) \right| 0\right\rangle\right].        \label{18}
\end{eqnarray}%
The $\gamma $ algebraic relation $\gamma _{\mu }\gamma _{\nu }\gamma _{5}=-
\frac{1}{2}\epsilon _{\mu \nu \alpha \beta }\sigma ^{\alpha \beta }+g_{\mu \nu }\gamma
_{5}$ can help us to write it down in a preferred form,
\begin{eqnarray}
G^{QCD}_{\mu }\left( p^2,\left( p+q\right)^2 \right) &=&\frac{i}{
\left( 2\pi \right) ^{4}}\int\int
d^{4}xd^{4}k\frac{1}{m_{b}^{2}-k^{2}} e^{i\left( p-k\right)
x}\left[ k_{\mu }\left\langle \rho \left| \overline{u} \left(
x\right) \gamma _{5}d\left( 0\right) \right| 0\right\rangle
\right.
\nonumber \\
&&\left.-\frac{1}{2}\epsilon_{\mu \nu \alpha \beta }k^\nu\left\langle \rho \left|
\overline{u}\left( x\right) \sigma ^{\alpha \beta }d\left( 0\right) \right| 0\right
\rangle +m_b\left\langle \rho \left| T\overline{u}\left( x\right) \gamma _{\mu }\gamma
_{5}d\left( 0\right) \right| 0\right\rangle\right] .  \label{19}
\end{eqnarray}%
The nonlocal matrix element $\left\langle \rho \left| \overline{u}\left( x\right)
\gamma _{5}d\left( 0\right) \right| 0\right\rangle $ is exactly vanishing, as required
by the parity conservation of strong interactions. The light cone expansions of the
other two matrix elements reads\cite{10,19} respectively,
\begin{eqnarray}
\left\langle \rho \left| \overline{u}\left( x\right) \sigma ^{\alpha \beta }d\left(
0\right) \right| 0\right\rangle
&=&-if_{\rho }^{T}\left\{ \left( e_{\left( \lambda
\right) }^{\ast \alpha }q^{\beta }-e_{\left( \lambda \right) }^{\ast \beta }q^{\alpha
}\right) \int_{0}^{1}due^{iuq\cdot x}\left[ \varphi _{\perp}\left( u\right)
+\frac{1}{16}m_{\rho }^{2}x^{2}A_{T}\left(u\right) \right] \right.  \nonumber \\
&&+\left( q^{\alpha }x^{\beta }-q^{\beta }x^{\alpha }\right) \frac{e_{\left( \lambda
\right) }^{\ast }\cdot x}{\left( q\cdot x\right) ^{2}}m_{\rho
}^{2}\int_{0}^{1}due^{iuq\cdot x}B_{T}\left( u\right)  \nonumber \\
&&\left.+\frac{1}{2}\left( e_{\left( \lambda \right) }^{\ast \alpha }x^{\beta
}-e_{\left( \lambda \right) }^{\ast \beta }x^{\alpha }\right) \frac{m_{\rho
}^{2}}{q\cdot x}\int_{0}^{1}due^{iuq\cdot x}C_{T}\left( u\right)\right\}, \label{20}
\end{eqnarray}
and
\begin{eqnarray}
\left\langle \rho \left| T\overline{u}\left( x\right) \gamma _{\mu }\gamma _{5}d\left(
0\right) \right| 0\right\rangle=\frac{1}{4}f_{\rho}m_{\rho}\varepsilon_{ \mu\alpha
\beta\gamma }q^{\alpha}e^{\beta}x^{\gamma}\int^1_0due^{iuq\cdot
x}g^{(a)}_{\perp}(u,\mu_{b}).       \label{21}
\end{eqnarray}
$\varphi _{\perp}\left( u\right) $  and $ g^{(a)}_{\perp}(u)$ stand for the twist-2
and -3 wavefunctions, respectively, and obey $\int_{0}^{1}f\left( u\right) du =1$; the
others are associated with twist-4 operators and parametrize the $\rho $ mass
corrections, among which both $ B_{T}\left( u\right) $ and $C_{T}\left( u\right) $
abide by $ \int_{0}^{1}f\left( u\right) du=0$ as it stands.

As with the $ g_{BB\rho }$ case, in practical calculations we use
the definitions $\overline{B} _{T}\left( u\right) =\int_{0}^{u}
B_{T}\left( v\right) dv$, $\widetilde{B} _{T}\left( u\right)
=\int_{0}^{u}\overline{B}_{T}\left( v\right) dv$ and $\overline{G}
_{T}\left( u\right) =\int_{0}^{u}G_{T}\left( v\right) dv$.
Substituting Eqs. (20) and (21) into (19) we gain the theoretical
expression for $\widetilde{G}^{QCD}(p^2,\left(p+q\right)^2)$,
\begin{eqnarray}
\widetilde{G}^{QCD}\left( p^2,\left( p+q\right)^2 \right)
&=&f_{\rho }^{T}\left\{ \int_{0}^{1}du \frac{\varphi
_{\perp}\left( u\right)}{m_{b}^{2}-\left( p+uq\right) ^{2}}
+\frac{1}{4}m_{\rho }^{2}\int_{0}^{1}duA_{T} (u)\right.  \nonumber \\
&&\left.\times \left[ \frac{3}{\left[ m_{b}^{2}-\left( p+uq\right) ^{2}\right] ^{2}}-
\frac{2m_{b}^{2}}{\left[ m_{b}^{2}-\left( p+uq\right) ^{2}\right] ^{3} }
\right]\right\} \nonumber\\
&&+\frac{1}{2}m_bm_{\rho}f_{\rho}\int^1_0du\frac{g^{(a)}_{\perp}(u)}{\left[
m_{b}^{2}-\left( p+uq\right) ^{2}\right] ^{2}}. \label{22}
\end{eqnarray}%
It is interesting to note that the $\rho $ mass effect is only offered by the twist-4
wavefunction $A_{T}(u)$. Obviously, the QCD spectral densities concerning the leading
and next-to-leading terms are, except for a constant factor, the same as those in the
$g_{BB\rho}$ case, respectively. Omitting details, we end up with the following sum
rule for the product $f_{B^ {\ast }}f_{B}g_{B^{\ast }B\rho }$,
\begin{eqnarray}
f_{B^{\ast }}f_{B}g_{B^{\ast }B\rho } &=&\frac{m_{b}}{ m_{B^{\ast
}}m_{B}^{2}}e^{\frac{m_{B^{\ast }}^2+m_{B}^{2}}{2{M}^{2}} }\left\{ f_{\rho
}^{T}\left[\left( e^{-\frac{1}{{M}^{2}}\left( m_{b}^{2}+\frac{1}{4} m_{\rho
}^{2}\right) }-e^{-\frac{s_0}{{M}^{2}}}\right)M^{2}\varphi _{\perp}\left(
\frac{1}{2}\right)\right.\right.
\nonumber \\
&&\left.+\frac{1}{4}m_{\rho }^{2}\left( 3-\frac{m_{b}^{2}}{{M}^{2}}\right) e^{-
\frac{1}{{M}^{2}}\left( m_{b}^{2}+\frac{1}{4}m_{\rho }^{2}\right) }
A_{T}\left( \frac{1}{2}\right)\right]  \nonumber\\
&&\left. +\frac{1}{2}f_{\rho}m_bm_{\rho}e^{-\frac{1}{2}\left( m_{b}^{2}+\frac{1}{4}
m_{\rho }^{2}\right)}g^{(a)}_{\perp}\left(\frac{1}{2}\right)\right\}. \label{23}
\end{eqnarray}
It should be understood that the Borel variables ${M}_{1}^{2}$ and ${M}_{2}^{2}$ have
been taken equal once again, for the $B$ and $B^{\ast }$ mesons are nearly degenerate
in mass. This enables the relevant wavefunctions to take values at $u=1/2$, to high
accuracy.

We would like to emphasize that the previous LCSR results for $g_{B^{\ast}B\rho}$ and
$g_{D^{\ast}D\rho}$\cite{13} are questionable, because of a missing factor ${1}/{2}$
in front of the term $-\epsilon_{\mu\alpha\lambda\rho}\sigma^{\lambda\rho}$ in writing
down the $\gamma$ algebraic relation $\gamma_{\mu}\gamma_{\alpha}
\gamma_{5}=-{1}/{2}\epsilon_{\mu\alpha\lambda\rho}\sigma^{\lambda\rho}+g_{\mu\alpha}$,
and a misuse of the subtraction procedure $e^{-\frac{1}{M^2}\left(m_b^2+\frac{1}{4}
m_{\rho}^2\right)}\rightarrow e^{-\frac{1}{M^2}\left (m_b^2+\frac{1}{4}m_{\rho}^2
\right)}-e^{-\frac{s_0}{M^2}}$ imposed on the twist-3 parts.
\begin{center}
\noindent{\large \bf 3. Numerical Results }
\end{center}

The parameters to need fixing for a numerical estimate are those concerning the $B^{
\ast }$\ or $B$ mesons and those describing the $\rho $ meson. The former contain the
decay constants $f_{B}$ and $f_{B^{\ast }}$, mass parameters $m_B$, $m_{B^*}$ and
$m_b$, and threshold parameter $s_{0}$. We use $m_b=4.8\pm0.1\gev$, $m_B=5.279\ GeV$
and $m_{B^*}=5.325\gev$. With the two point sum rules formulated in\cite{8}, the
values of the decay constants $f_{B}$ and $f_{B^{\ast}}$ are fixed at $f_{B}=115\mev$
and $f_{B^{\ast}}=125\mev$, corresponding to $m_b=4.8\gev,\ s_0=33\gev^2$ and the
leading order in $\alpha_s$. Also, all the parameters of the $\rho $ meson involved
become now numerically available. We take as inputs the experimental values $m_{\rho
}=770\ MeV$, $f_{\rho }=198\pm 7\ MeV$ and the QCD sum rule result $f_{\rho
}^{T}=152\pm 9\ MeV$ \cite{18} at the scale $u_b=\sqrt{m_{B}^{2}-m_{b}^{2}}\approx
2.5\ GeV.$ Concerning the light cone wavefunctions appearing in our sum rules, there
have been many discussions on them in the literature. It is in \cite{17} that QCD sum
rule method is first applied to study the twist-2 distribution amplitudes of vector
mesons. Later on, a more systematic discussion was given in \cite{18}, where results
of \cite{17} were critically examined and updated. Very recently, the authors of
\cite{19} took further the meson mass corrections into consideration by introducing
some higher twist distributions in the light cone expansions of the relevant nonlocal
matrix elements, extending the work \cite{18} by an additional use of the QCD
equations of motion. The yielded findings, some of which will be used in our numerical
analysis, have found applications\cite{10} in phenomenology of exclusive semileptonic
and radiative $B$ decays. The explicit forms of the light cone wavefunctions in
relation to our sum rule calculations are
\begin{eqnarray}
\varphi _{\perp }\left( u,\mu _{b}\right) \!\!\!&=&\!\!\!6u\left( 1-u\right) \left(
1+a_{2}^{\perp }\left( \mu _{b}\right) \frac{3}{2}\left[ 5\left( 2u-1\right)
^{2}-1\right] \right), \nonumber \\ \varphi _{\parallel }\left( u,\mu _{b}\right)
\!\!\!&=&\!\!\!6u\left( 1-u\right) \left( 1+a_{2}^{\parallel }\left( \mu _{b}\right)
\frac{3}{2}\left[ 5\left( 2u-1\right)
^{2}-1\right] \right),\nonumber\\
h^{(s)}_{\parallel }\left( u,\mu
_{b}\right)\!\!\!&=&\!\!\!6u(1-u)(1+0.15[5(2u-1)^2-1]),
\nonumber\\
g^{(a)}_{\perp }\left( u,\mu _{b}\right)\!\!\!&=&\!\!\!6u(1-u)(1+[5(2u-1)^2-1]),\label{24}\\
A_{T}\left( u,\mu _{b}\right) \!\!\!&=&\!\!\!24u^{2}\left( 1-u\right) ^{2}, \nonumber\\
A\left( u,\mu _{b}\right) \!\!\!&=&\!\!\!\left[ \frac{4}{5}+\frac{20}{9}\xi _{4}+\frac{%
8}{9}\xi _{3}\right] 30u^{2}\left( 1-u\right) ^{2},\nonumber\\
\widetilde{C}\left(u,\mu_b\right)\!\!\!&=&\!\!\!-\left[\frac{3}{2}+\frac{10}{3}\xi_4+\frac{10}{3}\xi_{3}
\right]u^2(1-u)^2, \nonumber
\end{eqnarray}
with the coefficient $a_{2}^{\perp }\left( \mu _{b}\right) =0.17\pm 0.09$ and
$a_{2}^{\parallel }\left( \mu _{b}\right) =0.16\pm 0.09$, $\xi_3=0.023$ and
$\xi_4=0.13$.

Having all the input parameters at hand, we could carry out the numerical
calculations. It is a critical step towards deriving a reliable sum rule prediction to
look for a reasonable range of the Borel parameters. The standard procedure requires
that the terms proportional to the highest inverse power of the Borel parameters stay
reasonably small, which can fix the lower limit of the fiducial Borel interval, and
that the higher resonance and continuum contribution should not become too large,
which may determine the upper limit of the allowed range. For the two sum rules in
consideration, we find that the Borel intervals to satisfy the above criteria are
respectively $8\leq M^{2}\leq 14\gev^{2}$ for the $g_{BB\rho}$ case and $8\leq
M^{2}\leq 15\gev^{2}$ for the $g_{B^{\ast}B\rho}$ case, where the twist-4
wavefunctions contribute less than $6\%$ and $5\%$ and the high states at the orders
lower than $23\%$ and $25\%$, respectively. The figure 1 shows the sensitivity of the
sum rules for $f_B^2g_{BB\rho}$ and $f_{B^{\ast}}f_Bg_{B^{\ast}B\rho}$ to the Borel
parameters. From the corresponding sum rule "windows", we arrive at
$f_{B}^{2}g_{BB\rho }=0.071\pm 0.002\ GeV^2$ and $f_{B^{\ast }}f_{B}g_{B^{\ast }B\rho
}=0.082\pm 0.004\ GeV$, the uncertainties quoted being due to the variations of
$M^{2}$. By means of values of decay constants obtained previously it is immediate to
get the desired sum rules for the strong couplings $g_{BB\rho }$ and $ g_{B^{\ast
}B\rho }$. If using the central values for all the relevant sum rule results, we have
$g_{BB\rho}=5.37$ and $g_{B^*B\rho}=5.70\ GeV^{-1}$.

For a better understanding of the overall uncertainties in the coupling constants, it
is highly advisable to employ the analytic forms instead of the numerical results in
Eqs.(17) and (23) for the decay constants. When $m_{b}$ keeps fixed while $s_{0}$
changes between $32-34\ GeV^{2}$, the resulting variations relative to the central
values amount to $\pm 6\%$ for $g_{BB\rho }$\ and to $\pm 8\%$ for $g_{B^{\ast }B\rho
}$, if the uncertainties due to the Borel parameters are included. The influences on
the sum rules can be investigated of the uncertainty in $ m_{b}$, by considering a
correlated variation of $m_{b}$ and $s_{0}$ in the individually allowed ranges.
Requiring the strong couplings $g_{BB\rho }$ and $g_{B^{\ast }B\rho }$ to take values
only if the sum rules for the relevant decay constants show the best stability, we
observe that the induced changes are typically of orders $7\%$ and $6\%$ respectively.
It is of course important to investigate further the uncertainties from the
wavefunctions. The simplest way to test the sensitivity of the sum rules to model
wavefunctions is by putting all the corresponding nonasymptotic coefficients to zero,
namely by using their asymptotic forms which are mode-independent and completely
dictated by perturbative QCD. The resulting sum rules deviate from their individual
central values by about $7\%$ in the $g_{BB\rho }$ case and by about $8\%$ in the
$g_{B^{\ast }B\rho }$ case. The effects observed in such a way come from an extreme
treatment and therefore are anyway being overestimated. Taking it into account that
the twist-4 wavefunctions bring only a correction of about $4\%$ to both sum rules for
$g_{BB\rho }$ and $g_{B^{\ast }B\rho },$ we can reasonably conjecture that the
uncertainties due to neglected yet higher twists would be at best of the same orders
as the twist-4 corrections. At present, the total uncertainties in the sum rules for
$g_{BB\rho}$ and $g_{B^*B\rho}$ can conservatively be estimated to be about $25\%$ and
$27\%$, respectively, by adding linearly up all the considered errors.

The same procedures may be used for a numerical discussion of LCSR for $g_{DD\rho}$
and $g_{D^{\ast}D\rho}$. The relevant parameters are taken as $m_c=1.3\gev$,
$m_D=1.87\gev$, $m_{D^{\ast}}=2.01\gev$, $f_D=170\mev$, $f_{D^{\ast}}=240\mev$ and
$s_0=6\gev^2$. In addition, we have to evolute the wavefunctions to a lower scale
$\mu_c=\surd\overline{m_D^2-m_c^2}$. Using the standard criteria the fiducial
intervals of $M^2$ turns out to be $4<M^2<8\gev^2$ in the $g_{DD\rho}$ case and
$5<M^2<8\gev^2$ in the $g_{D^{\ast}D\rho}$ case. The stability of the sum rules for
$f_D^2g_{DD\rho}$ and $f_{D^{\ast}}f_Dg_{D^{\ast}D\rho}$ is illustrated in Fig. 2. We
have $f_D^2g_{DD\rho}=0.11\gev^2$ and $f_{D^{\ast}}f_Dg_{D^{\ast}D\rho} =0.17\gev$,
with the negligibly small uncertainties due to $M^2$. As for the strong couplings
$g_{DD\rho}$ and $g_{D^{\ast}D\rho}$, the resulting sum rules are predicted to be
$g_{DD\rho}=3.81$ and $g_{D^{\ast}D\rho}= 4.17\gev^{-1}$, the total uncertainties
being about $23\%$ and $25\%$, respectively.

\begin{center}
\noindent{\large\bf 4. SUMMARY}
\end{center}

We have made an intensive study on QCD interactions between heavy
mesons and a light vector meson within the framework of LCSR. A
detailed deviation of the sum rules is presented for the relevant
strong coupling constants $g_{BB\rho }$($g_{DD\rho}$) and
$g_{B^{\ast }B\rho }$($g_{D^{\ast}D\rho}$) and a systematic
numerical analysis, including a painstaking investigation of the
uncertainty arising from all the possible sources of error, is
made. An existing negligence is pointed out in the previous LCSR
calculation on $g_{B^{\ast}B\rho}$ and $g_{D^{\ast}D\rho}$, and an
updated LCSR result is formulated.

The obtained predictions can be used to estimate the couplings for the other charge
states using the relations from isospin symmetry. Also, it is straightforward to
investigate the $B_sBK^{\ast}$, $B_s^*BK^{\ast}$ and $B^*B_s K^{\ast}$ strong
couplings, and the corresponding those in $c$ quark meson case by making a
corresponding parameter replacement in the relevant sum rules formulated.

The numerical results presented here should be updated, once our understanding of the
meson wavefunctions, $b$-quark mass and decay constants became more clear, and the QCD
radiative corrections are included in sum rule calculation.

\newpage
\begin{figure}
\centerline{ \epsfxsize=12cm \epsfbox{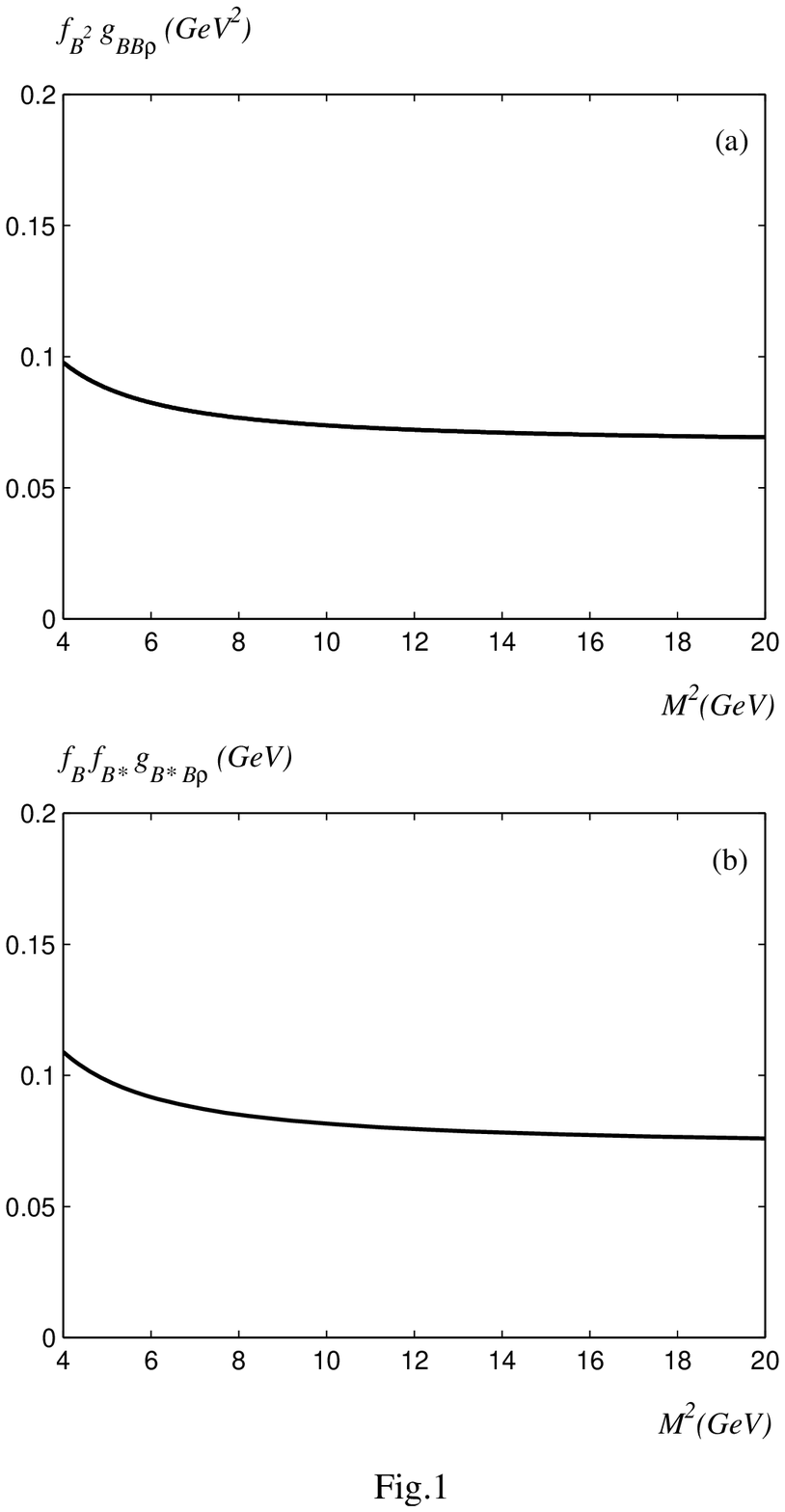} }
\end{figure}
\begin{figure}
\centerline{ \epsfxsize=12cm \epsfbox{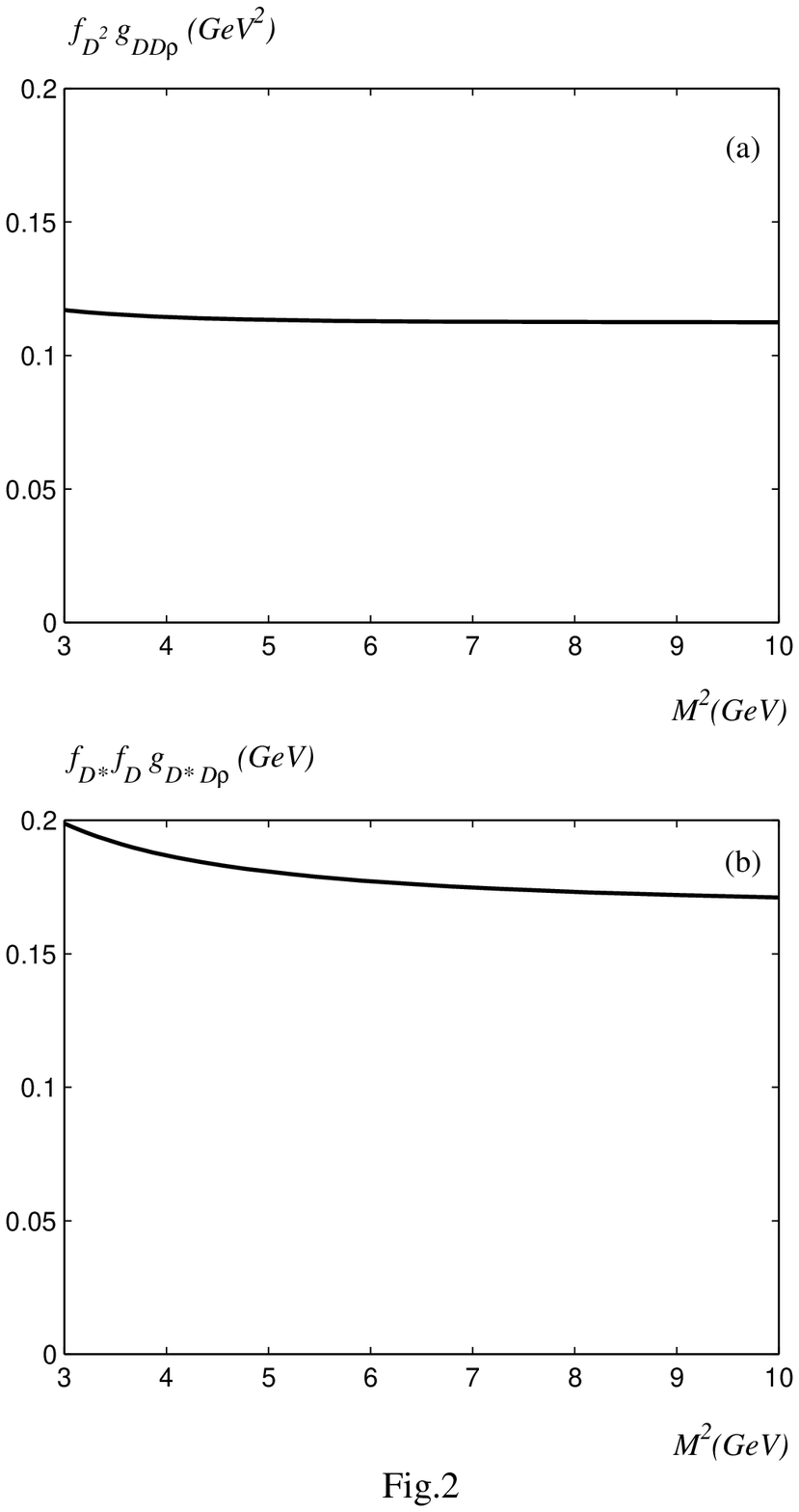} }
\end{figure}
\newpage
\begin{center}{\large \bf FIGURE CAPTIONS}
\end{center}
\ \\
\noindent Fig.1: The stability of LCSR for the products $f_B^2g_{BB\rho}$(Fig.1 (a))
and $f_{B^{\ast}}f_Bg_{B^{\ast}B\rho }$(Fig.1 (b)), with $m_b=4.8\ GeV$ and $s_0=33\
GeV^2 $.\\
\noindent Fig.2: The stability of LCSR for the products $f_{D^2}g_{DD\rho}$(Fig.2(a))
and $f_{D^{\ast}}f_Dg_{D^{\ast}D\rho}$(Fig.2(b)), with $m_c=1.3\gev$ and
$s_0=6\gev^2$.

\end{document}